\documentclass[prl,twocolumn,showpacs,preprintnumbers,amsmath,amssymb]{revtex4}

\usepackage{graphicx}
\usepackage{dcolumn}
\usepackage{bm}

\newcommand{\non}{\nonumber}
\newcommand{\rr}{{\bm r}}
\newcommand{\cdd}{c_{\rm dd}}
\newcommand{\RTF}{R_{\rm TF}}
\newcommand{\aho}{a_{\rm ho}}
\newcommand{\xid}{\xi_{\rm dd}}
\newcommand{\xisp}{\xi_{\rm sp}}
\newcommand{\ee}{{\bm e}}

\begin{document}

\preprint{APS/123-QED}


\title{Spontaneous Circulation in Ground-State Spinor Dipolar Bose-Einstein Condensates}
\author{Yuki Kawaguchi$^1$}
\author{Hiroki Saito$^2$}
\author{Masahito Ueda$^{1,3}$}
\affiliation{$^1$Department of Physics, Tokyo Institute of Technology,
2-12-1 Ookayama, Meguro-ku, Tokyo 152-8551, Japan \\
$^2$The University of Electro-Communications,
1-5-1, Choufugaoka, Choufu-shi, Tokyo 182-8585, Japan \\
$^3$Macroscopic Quantum Control Project, ERATO, JST, Bunkyo-ku, Tokyo 113-8656, Japan
}
\date{\today}

\begin{abstract}
We report on a study of the spin-1 ferromagnetic Bose-Einstein condensate with magnetic dipole-dipole interactions.
By solving the non-local Gross-Pitaevskii equations for this system,
we find three ground-state phases.
Moreover, we show that a substantial orbital angular momentum accompanied by chiral symmetry breaking emerges 
spontaneously in a certain parameter regime.
We predict that all these phases can be observed in the spin-1 $^{87}$Rb condensate by changing the number of atoms or the trap frequency.
\end{abstract}


\maketitle

The magnetic dipole-dipole interaction in ferromagnets
is responsible for a rich variety of spin structures~\cite{DomainWalls}.
Similar spin textures can also be expected to occur,
due to the dipolar interaction,
in ferromagnetic Bose-Einstein condensates (BECs), such as the spin-1 $^{87}$Rb BEC.
However, spinor BECs differ from ferromagnets in that
they exhibit spin--gauge symmetry that can generate mass flow by developing spin textures~\cite{Ho1996}.
One may therefore wonder whether the dipole-induced spin texture can yield spontaneous mass current in the ground state.
In this Letter we show that this is the case.

An additional motivation for our work is the recent observation of
a dipolar BEC in a system of spin-polarized $^{52}$Cr atoms~\cite{CrBEC}.
The ground state of spinor dipolar BECs has been studied by several researchers using a single-mode approximation~\cite{Yi2004,Santos2006,Diener2006}.
Magnetism of dipolar BECs in one and two dimensional optical lattices has also been investigated~\cite{Pu2001,Gross2002}.
In this Letter, we study the ground-state spin textures of a spinor dipolar BEC without invoking the single-mode approximation.
We show the existence of three ground-state phases at zero magnetic field.
In particular, we identify chiral spin-vortex phase
and show that spontaneous circulation with broken chiral symmetry emerges in this phase.

We consider a system of $N$ spin-1 atoms with mass $M$
confined in a spin-independent potential $U_{\rm trap}(\rr)=M\omega^2(x^2+y^2+z^2)/2$.
The Hamiltonian of the system is given by~\cite{SpinorBEC}
\begin{align}
 \hat{H}_{\rm tot}&=\int d\rr \left[
 \hat{\psi}_{m_1}^\dagger(\rr) H_0 \hat{\psi}_{m_1}(\rr) \right.\non\\
 &+ \frac{g_0}{2} \hat{\psi}_{m_1}^\dagger(\rr)\hat{\psi}_{m_2}^\dagger(\rr) \hat{\psi}_{m_2}(\rr)\hat{\psi}_{m_1}(\rr) \non\\
 &+ \left. \frac{g_1}{2} \hat{\psi}_{m_1}^\dagger(\rr)\hat{\psi}_{m_2}^\dagger(\rr)
 {\bm F}_{m_1m_4}\cdot{\bm F}_{m_2m_3} \hat{\psi}_{m_3}(\rr)\hat{\psi}_{m_4}(\rr)
 \right]\non\\
 & + \hat{H}_{\rm dd},
\end{align}
where $\hat{\psi}_m(\rr)$ is the annihilation operator of an atom in the magnetic sublevel $m=0,\pm1$ at point $\rr$,
$H_0=-\hbar^2\nabla^2/(2M)+U_{\rm trap}(\rr)$, 
${\bm F}_{mm'}$ represents the spin-1 matrix,
and $\hat{H}_{\rm dd}$ is the contribution of the magnetic dipole-dipole interaction.
The spin-independent and spin-dependent interactions are characterized by
$g_0=4\pi\hbar^2(a_0+2a_2)/(3M)$ and $g_1=4\pi\hbar^2(a_2-a_0)/(3M)$, respectively,
with $a_s$ $(s=0,2)$ being the {\it s}-wave scattering length for the scattering channel with total spin $s$.
For spin-1 $^{87}$Rb atoms
we have $0< -g_1 \ll g_0$, and the ground state is ferromagnetic.

The magnetic dipole moment of an atom is given by
$\hat{\bm \mu} = -g_S\mu_{\rm B} \hat{\bm S} + g_I\mu_{\rm N}\hat{\bm I}$,
where $\mu_{\rm B}$ is the Bohr magneton, $\mu_{\rm N}$ is the nuclear magneton,
$\hat{\bm S}$ and $\hat{\bm I}$ are the electronic and nuclear spin angular momenta, respectively, 
and $g_S$ and $g_I$ are the Land\'{e} {\it g} factors.
The matrix element of the dipole moment between magnetic sublevels in the spin-1 hyperfine manifold is shown to be
$\langle m|\hat{\bm \mu}|m'\rangle = g_F \mu_{\rm B} {\bm F}_{mm'}$.
For the case of $^{87}$Rb with $S=1/2$ and $I=3/2$,
we have $g_F=(g_S+5g_I\mu_{\rm N}/\mu_{\rm B})/4\simeq 1/2$.
Then the contribution of the magnetic dipole-dipole interaction to the Hamiltonian is given by
\begin{align}
\hat{H}_{\rm dd} &= \frac{\cdd}{2}\int d\rr \int d\rr'
\hat{\psi}_{m_1}^\dagger(\rr)\hat{\psi}_{m_2}^\dagger(\rr')\hat{\psi}_{m_3}(\rr')\hat{\psi}_{m_4}(\rr)\non\\
&\times\frac{{\bm F}_{m_1m_4}\cdot{\bm F}_{m_2m_3} - 3({\bm F}_{m_1m_4}\cdot\ee)({\bm F}_{m_2m_3}\cdot\ee)}{|\rr-\rr'|^3},
\end{align}
where $\cdd=\mu_0g_F^2\mu_{\rm B}^2/(4\pi)$ with
$\mu_0$ being the magnetic permeability of the vacuum,
and $\ee=(\rr-\rr')/|\rr-\rr'|$.
We ignore the coupling of the $F=1$ manifold with the $F=2$ manifold due to the dipolar interaction
because the hyperfine splitting $\sim 100$ mK is much larger than the dipolar energy $\sim 1$ nK.

Our system has three characteristic length scales:
dipole healing length $\xid = \hbar/\sqrt{2M\cdd n_0}$,
spin healing length $\xisp = \hbar/\sqrt{2M|g_1| n_0}$,
and Thomas-Fermi (TF) radius $\RTF=2\aho [g_0n_0/(\hbar\omega)]^{1/2}$,
where $n_0=\hbar\omega/g_0[5(a_2+2a_0)N/(8\aho)]^{2/5}$ is the TF peak density
and $\aho=[\hbar/(2M\omega)]^{1/2}$.
As shown below, the phase diagram is characterized by two dimensionless parameters
$\RTF/\xid$ and $\RTF/\xisp$.

We investigate the mean-field ground state of the total Hamiltonian $\hat{H}_{\rm tot}$.
The general form of the ferromagnetic spinor order parameter is given by
\begin{align}
\begin{pmatrix} \psi_1\\ \psi_0\\ \psi_{-1}\end{pmatrix}
= \sqrt{n}e^{-i\gamma}
\begin{pmatrix}
e^{-i\alpha}  \cos^2(\beta/2)\\
\sqrt{2}\sin(\beta/2)\cos(\beta/2)\\
e^{i\alpha}  \sin^2(\beta/2)
\end{pmatrix},
\label{eq:OP}
\end{align}
where $n$ is the number density, $\gamma$ is the gauge,
and ${\bm s}=(\sin\beta\cos\alpha,\sin\beta\sin\alpha,\cos\beta)$ is the unit vector parallel to the magnetization.
The dipolar interaction induces a spin texture characterized by a spatial dependence of Euler angles $\alpha, \beta, \gamma$.
This dependence induces mass and spin velocity fields which can be calculated from
continuity equations
$\partial n/\partial t+\nabla\cdot(n{\bm v})=0$ and
$\partial (ns_\mu)/\partial t+\nabla\cdot(n{\bm v}_{\rm sp}^\mu + ns_\mu {\bm v})=0$, as
${\bm v}=-\hbar/M(\nabla\gamma+\cos\beta\nabla\alpha)$,
${\bm v}_{\rm sp}^z=-\hbar/(2M)\sin^2\beta\nabla\alpha$, and
${\bm v}_{\rm sp}^x+i{\bm v}_{\rm sp}^y=-\hbar/(2M)e^{i\alpha}\left[-\sin\beta\cos\beta\nabla\alpha+i\nabla\beta\right]$.
Here $n{\bm v}$ and $n{\bm v}_{\rm sp}^\mu$ $(\mu=x,y,z)$ describe the current density of the mass
and that of the $\mu$-component of the spin vector ${\bm s}$, respectively.
In addition to the kinetic energy $K_0=\int d\rr \hbar^2/(2M)|\nabla\sqrt{n}|^2$ arising from the spatial dependence of the total density,
the system has kinetic energies
$K_{\rm mass}=\int d\rr \frac{1}{2}nM|{\bm v}|^2$ and $K_{\rm spin}=\int d\rr nM\sum_\mu|{\bm v}_{\rm sp}^\mu|^2 = \int d\rr \hbar^2/(4M)\sum_\mu|\nabla s_\mu|^2$
caused by the mass and spin currents, respectively.
In the case of ferromagnets,
the domain structure is mainly determined by the interplay between the dipolar interaction
and the spin stiffness that has the form of $K_{\rm spin}$.
In the case of dipolar BECs, the additional term $K_{\rm mass}$ is significant.

The ground-state phases can be classified by the symmetry of the order parameter.
The total Hamiltonian $\hat{H}_{\rm tot}$ is invariant under
space inversion ${\sf P}\hat{\psi}_m(\rr)=\hat{\psi}_m(-\rr)$
and time reversal ${\sf T}\hat{\psi}_m(\rr)=(-1)^m\hat{\psi}_{-m}^\dagger(\rr)$.
It also possesses rotational [i.e., SO(3)] symmetry in combined spin and coordinate space around an arbitrary axis.
However, as we show below,
the SO(3) symmetry is reduced to uniaxial symmetry
and the obtained phases are eigenstates of the projected total angular momentum $\hat{J}_z=-i(\partial/\partial\varphi)+F_z$
in cylindrical coordinates $\rr=(r,\varphi,z)$, characterized by
\begin{subequations}
\begin{align}
 \alpha(\rr) &= \varphi + \widetilde{\alpha}(r,z),\\
 \beta(\rr)  &= \beta(r,z) = \beta(r,-z),\\
 \gamma(\rr) &= -J\alpha(\rr) + \widetilde{\gamma}(r,z),
\end{align}
\end{subequations}
with $J$ being the eigenvalue of $\hat{J}_z$.
We thus classify the obtained phases by the value of $J$,
space-inversion symmetry, and time-reversal symmetry.

The ground-state phases are obtained by solving the non-local Gross-Pitaevskii (GP) equations,
\begin{subequations}
\begin{align}
(H_0 + g_0n)\psi_0 +(g_1f_+ - \cdd b_+)\frac{\psi_1}{\sqrt{2}}& \non\\
+(g_1f_- - \cdd b_-)\frac{\psi_{-1}}{\sqrt{2}}& = \mu\psi_0, \\
(H_0 + g_0n)\psi_{\pm1} + (g_1 f_{\mp} - \cdd b_{\mp})\frac{\psi_0}{\sqrt{2}}& \non\\
 \pm (g_1 f_z - \cdd b_z)\psi_{\pm1}& = \mu\psi_{\pm1},
\end{align}
\label{eq:GP}\end{subequations}
where  $\mu$ is the chemical potential, $n=\sum_m|\psi_m|^2$ is the number density,
and ${\bm f}=(f_x,f_y,f_z)$ is the spin density
defined by $f_z=|\psi_1|^2-|\psi_{-1}|^2$ and $f_x+if_y=f_+=f_-^*=\sqrt{2}(\psi_1^*\psi_0+\psi_0^*\psi_{-1})$.
When the system is purely ferromagnetic, the spin density satisfies ${\bm f}=n{\bm s}$.
We define the scaled dipole field ${\bm b}=(b_x,b_y,b_z)$ as
$b_\mu(\rr) = - \sum_{\nu=x,y,z}\int d\rr' (\delta_{\mu\nu}-3e_\mu e_\nu)f_\nu(\rr')/|\rr-\rr'|^3$
and $b_{\pm}=b_x\pm i b_y$.
The effective magnetic field produced by the surrounding magnetic dipoles is given by $\cdd {\bm b}/(g_F\mu_{\rm B})$.

We numerically solve Eq.~\eqref{eq:GP} in three dimensions
by the imaginary-time propagation method.
We prepare the initial state by first solving Eq.~\eqref{eq:GP} with $\cdd=0$
and then randomizing the local spin directions while keeping the number density unchanged.
Comparing the total energies of the obtained textures,
we obtain the phase diagram in the parameter space of $(\RTF/\xisp,\RTF/\xid)$
as shown in Fig.~1(a).
For the spin-1 $^{87}$Rb BEC, we have $\xisp/\xid=\sqrt{\cdd/|g_1|}=0.30$,
and the system follows the dotted line in Fig.~1(a).
A typical spin configuration in each phase is shown in Figs.~1(b)--1(d).
The symmetry properties are summarized in Fig.~2.
We now describe the properties of each phase.

\begin{figure} 
\includegraphics[width=0.95\linewidth]{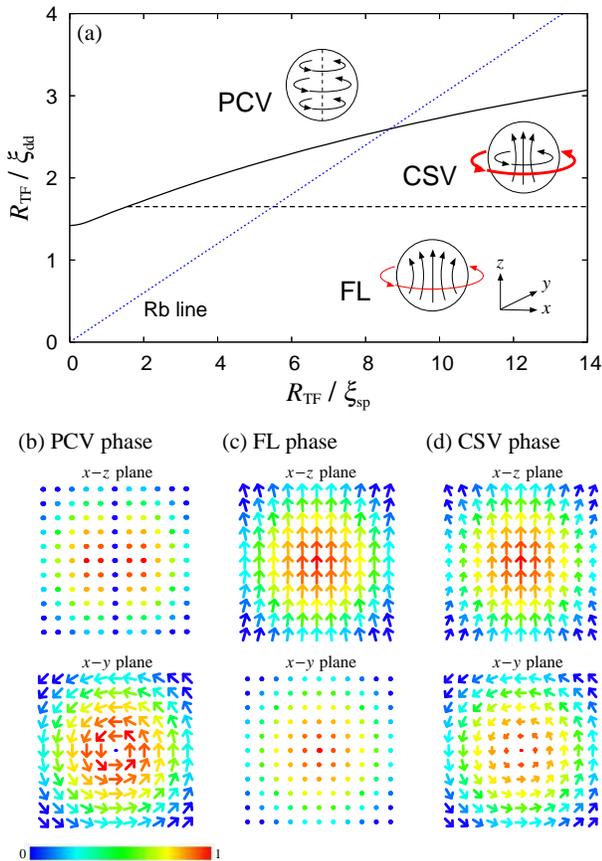}
\caption{(Color) (a) Phase diagram of a ferromagnetic dipolar BEC.
The solid curve shows the first-order phase boundary between the $J=0$ and $J=1$ phases
and the broken line represents the second-order one and divides the phases with and without spin chirality,
where PCV, CSV, and FL stand for polar-core vortex, chiral spin-vortex, and flower phases, respectively.
The schematic shown in each region represents the spin configuration (black arrows) and mass flow (red arrows).
The spin-1 $^{87}$Rb BEC ($\xisp/\xid=0.30$) traces the dotted line, which is referred to as the Rb line, as a function of $\RTF/\xisp$.
(b)--(d) Typical spin configuration in each phase.
The top and bottom panels show the unit vector ${\bm f(\rr)}/|{\bm f}(\rr)|$ projected onto the $x$--$z$ and $x$--$y$ planes, respectively.
The color of the arrows represents the magnitude of the spin density normalized by the number density $|{\bm f}(\rr)|/n(\rr)$ (see the bottom scale).
}
\end{figure}
\begin{figure}
\includegraphics[width=0.95\linewidth]{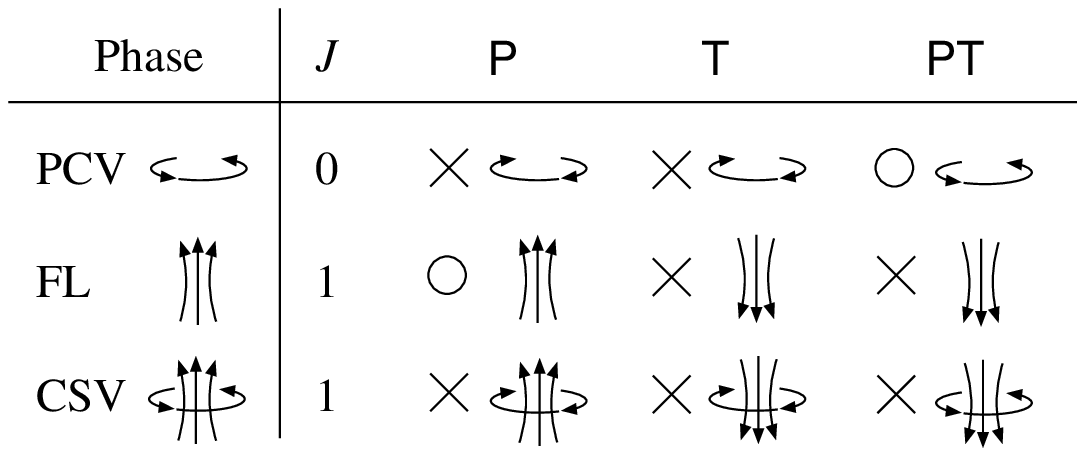}
\caption{Broken ($\times$) and unbroken ($\bigcirc$) symmetries for each phase in Fig.~1.
The schematic in each cell shows the spin configuration after the transformation by {\sf P}, {\sf T}, and {\sf PT}.
}
\end{figure}

{\it Polar-core vortex (PCV) phase}.---%
This phase has total angular momentum $J=0$ and
has a polar-core, i.e., $(\psi_1,\psi_0,\psi_{-1})=\sqrt{n}(0,1,0)$ at $r=0$~\cite{Isoshima2001}.
The length of the spin vector $|{\bm f}|/n$ is less than unity for $r\lesssim\xisp$,
and the spin configuration far from the core is described by 
$\widetilde{\alpha}=\pm\pi/2$, $\beta=\pi/2$, and $\widetilde{\gamma}={\rm const.}$
Since the $m=1$ component has the phase winding number $-1$ and the $m=-1$ component has the phase winding number $+1$,
the net mass current vanishes while the net spin current remains, giving a nonzero $K_{\rm spin}$.
Time-reversal symmetry and space-inversion symmetry are both broken,
but the order parameter remains invariant under the combined {\sf PT} operation.

{\it Flower (FL) phase}~\cite{Schabes1988}.---%
This phase has $J=1$ and can continuously develop from a spin-polarized state.
The spin configuration is given by $\widetilde{\alpha}(z>0)=0$, $\widetilde{\alpha}(z<0)=\pi$, $0\le \beta \ll \pi/2$,
and $\widetilde{\gamma}={\rm const.}$
Since $\beta\sim0$, most of the total angular momentum $N\hbar$ resides in the spin
and the remaining small angular momentum is carried by mass current.
The spin current is also small.
The system has space-inversion symmetry while time-reversal symmetry is broken.

{\it Chiral spin--vortex (CSV) phase}.---%
This phase has $J=1$ and the spin structure is described by
$0\le\widetilde{\alpha}\le\pi$, $\widetilde{\alpha}(r,z)+\widetilde{\alpha}(r,-z)=\pi$, and $0\le\beta\le\pi$.
The gauge $\widetilde{\gamma}$ also varies in space.
As shown in Fig.~1(d), this phase has a coreless vortex whose size is of the order of $\xid$.
The system has substantial spin and mass currents around the rotational axis
and has nonzero $K_{\rm mass}$ and $K_{\rm spin}$.
Both time-reversal symmetry and space-inversion symmetry are broken.

\begin{figure}
\includegraphics[width=0.95\linewidth]{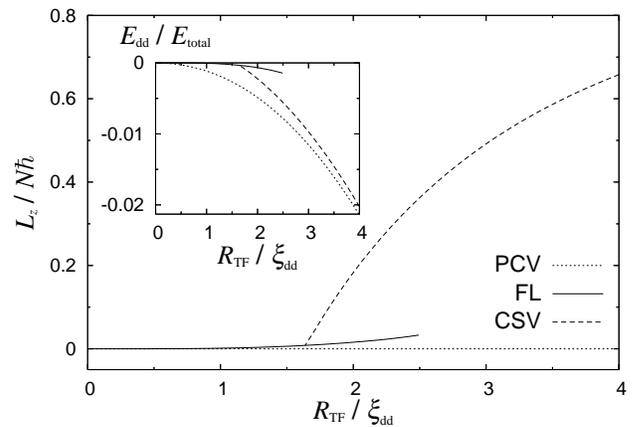}
\caption{Orbital angular momentum in each state as a function of $\RTF/\xid$.
The inset shows the ratio of the dipolar energy to the total energy
$E_{\rm dd}/E_{\rm tot}\equiv\langle \hat{H}_{\rm dd}\rangle/\langle \hat{H}_{\rm tot}\rangle$
for each state at $\RTF/\xisp=5.4$.
}
\label{fig:Lz}\end{figure}

The dependence of the dipolar energy is shown in the inset of Fig.~\ref{fig:Lz}.
The dipolar energy is minimized when the spin texture has a flux-closure structure~\cite{Landau1935},
i.e., when the divergence of the magnetization is everywhere zero.
The PCV state, which satisfies $\nabla\cdot{\bm f}=0$ in all space,
has the lowest dipolar energy of the three states.

The CSV state is different from the FL state in that the chiral symmetry of the spin vortex is
broken to form a flux-closure structure.
When the CSV state has a nonzero total angular momentum in the $+\hat{z}$ direction,
there are two degenerate whirling patterns of spins in the $x$--$y$ plane:
clockwise ($\widetilde{\alpha}<0$) and anticlockwise ($\widetilde{\alpha}>0$).
These patterns have the spin configurations of right-handed and left-handed screws,
which are transformed into each other under space inversion {\sf P}.
The system selects one of the degenerate states by breaking the chiral symmetry.
Accompanied by the chiral symmetry breaking, the net orbital angular momentum
$L_z=\int d\rr \sum_m \psi_m^*(-i\hbar\partial/\partial\varphi)\psi_m$
increases drastically as shown in Fig.~\ref{fig:Lz},
where $L_z$ in each state is plotted as a function of $\RTF/\xid$.
The FL and CSV states both have a finite orbital angular momentum,
but the dependence on $\RTF/\xid$ is quite different.
Since the order parameters in the FL and CSV states are purely ferromagnetic,
the spin texture depends only on $\RTF/\xid$, and therefore $L_z$ is independent of $\xisp$.

\begin{figure}
\includegraphics[width=0.9\linewidth]{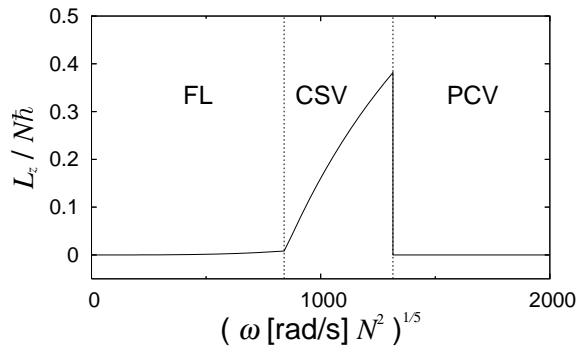}
\caption{
Orbital angular momentum as a function of $(\omega N^2)^{1/5}$ for $^{87}$Rb along the Rb line in Fig.~1.
}
\label{fig:exp}\end{figure}

Finally, we discuss possible experimental situations.
To form a spin texture in the ground state, the Zeeman energy must be smaller than the dipolar interaction energy,
otherwise the ground state is spin-polarized.
This implies that $B<10^{-5}$ G for the spin-1 $^{87}$Rb BEC with $n=10^{15}$ cm$^{-3}$.
The critical field increases in proportion to the atomic density.
By changing the number of atoms or the trap frequency,
all three phases can be experimentally realized.
Figure~\ref{fig:exp} shows the orbital angular momentum as a function of $(\omega N^2)^{1/5}$ along the Rb line in Fig.~1(a).
In the case of a BEC with $N=10^6$, for example, 
the CSV phase exists in the region of $2\pi\times 70$ Hz $\le \omega \le 2\pi\times 630$ Hz,
where $L_z$ increases up to $0.4N\hbar$.
It is interesting to note that we can induce a phase transition by controlling the trap frequency.

In summary,
we have investigated the phase diagram of a spin-1 ferromagnetic Bose-Einstein condensate with magnetic dipole-dipole interactions
and found three phases, namely, the polar-core vortex phase, the flower phase, and the chiral spin-vortex phases.
The chiral spin-vortex phase has chirality in the formation of the spin vortex,
and the topological spin structure spontaneously yields a substantial net orbital angular momentum.

\begin{acknowledgments}
This work was supported by Grant-in-Aids for Scientific Research (Grant
No.\ 17071005 and No.\ 17740263) and by 21st Century COE
programs on ``Nanometer-Scale Quantum Physics'' and ``Coherent Optical Science''
from the Ministry of Education, Culture, Sports, Science and Technology of Japan.
YK acknowledges support by the Japan Society for 
Promotion of Science (Project No.\ 185451).
MU acknowledges support by a CREST program of the JST.
\end{acknowledgments}

{\it Note added.}
--- The ground-state phase diagram of a spinor dipolar system has recently been discussed in Ref.~\cite{Yi2006}.



\end{document}